\documentclass[10pt]{article}
\usepackage{amsmath}
\usepackage{amssymb}
\usepackage{cite}
\usepackage{graphicx}
\usepackage{xcolor}
\usepackage{float}
\usepackage{caption}
\newcommand{\bey}{\begin{eqnarray}}
\newcommand{\eey}{\end{eqnarray}}
\usepackage[hmargin=2.9cm,vmargin=3.2cm]{geometry}

\begin{document}

\title{Thermodynamics of spherically symmetric thin-shell spacetimes}

\author{Demetrios Kotopoulis\thanks{d.kotopoulis@upatras.gr} and Charis Anastopoulos\thanks{anastop@upatras.gr} \\   Department of Physics, University of Patras, 26500 Greece}
\date{\today}

\maketitle

\begin{abstract}
We analyze the thermodynamics of spherically symmetric thin-shell solutions to Einstein's equations, including solutions with negative interior mass. We show the inclusion of such solutions is essential for the thermodynamic consistency of the system: the Maximum Energy Principle applies when we include an entropy term from the singularity of the negative-mass solutions, in addition to the Bekenstein-Hawking term for the entropy of solutions with positive interior mass. Then, the thermodynamic analysis leads to four distinct thermodynamic phases. We also show that all types of solutions can be either thermodynamically stable or dynamically stable, but only solutions with zero interior mass can be both. Since most of our results are analytic, thin shell models emerge as a useful theoretical paradigm for exploring  gravitational thermodynamics. Our results provide an additional argument in support of the assignment of entropy to the singularity of negative-mass Schwarzschild spacetimes, and, consequently, to Penrose's conjecture about the assignment of entropy to singularities.
\end{abstract}

\section{Introduction}

The characterization of black holes as thermodynamic objects by Bekenstein \cite{Bek1}  and Hawking  \cite{Hawk} has raised many foundational questions on the relation between gravity and thermodynamics \cite{wald01,pad10, Carlip}. These include the possibility of defining gravitational entropy in spacetimes that do not involve horizons, and the incorporation of both gravitational and  matter entropy into a consistent theoretical framework, as it is suggested by  the generalized second law of thermodynamics (GSL) \cite{Bek2}.

In this paper, we address these issues in one of the simplest gravitational systems, a spherically symmetric gravitating thin shell  \cite{Isra66, BI91, MS93, KMM06, Gon02}.
Thin shells provide elementary models  for many aspects of gravitational physics including  gravitational collapse \cite{lind74, gonc02, dels14, roch15},      quantum gravity \cite{haji00, vaz22}, and  black hole thermodynamics \cite{DFP86, Hiscock, lemos15}.  
Here, we analyse the thermodynamical properties of the most general static thin-shell solutions. These include solutions in which the interior of the shell is a Schwarzschild spacetime with negative mass, that is characterized by a naked singularity at the center. 
We find that the thermodynamics of the thin shell are consistent only if we include into the total entropy a specific contribution from the singularity that is determined uniquely. The resulting thermodynamic system admits different thermodynamic phases that correspond to positive, zero or negative mass of the solution inside the shell.

There are three distinct motivations for this work. The first one is Penrose's proposal that spacetime singularities may be carriers of entropy. In Refs. \cite{AnSav12, AnKot}, it was shown that self-gravitating radiation in a spherical box has a consistent thermodynamic description only if one assigns a specific value of entropy to the singular solutions to the Tolman-Oppenheimer-Volkoff equations. This result was later generalized for a box with an internal boundary, i.e., for a thick shell of radiation. These singular solutions are not specific to radiation, they are generic to Tolman-Oppenheimer-Volkoff  equations for any thermodynamically consistent equation of state (EoS) \cite{AnSav21}. The singularities are locally isometric to the singularities of Schwarzschild spacetime with negative mass, and they are relatively benign, as the spacetime is bounded-acceleration complete.  Thin shells provide a simpler set-up for a thermodynamic analysis, allowing for analytic proofs and for a precise derivation of the entropy terms for singularities.

Second, the thin-shell system also provides a simple model for   phase transitions between black holes and self-gravitating systems.  The most well known case of such a transition is the Hawking-Page phase transition between black holes and radiation \cite{HaPa}, albeit in asymptotically anti de Sitter spacetimes. In asymptotically flat spacetimes,  phase transitions have been studied by comparing the entropy of a Schwarzschild black hole in a box with the entropy of  radiation in the box \cite{Davies, Hut, Hawk76, Page2, York, AnSav16}. However, with the exception of Ref. \cite{AnSav16},  the matter degrees of freedom in these analyses are non-gravitating, i.e., they do not correspond to solutions of Einstein's equations. In contrast, the thin-shell models considered here are genuine solutions to Einstein's equations. Furthermore, they can incorporate a variety of different behaviors by changing the EoS, and they are analytically tractable.

Our third motivation   is the  physics of non-extensive thermodynamic systems in equilibrium. Non-extensivity appears when the range of interacting forces is larger than the size of the system. This is possible either in small systems with short-range forces, or in presence of long-range forces \cite{DRAW,CDR09}, such as gravity \cite{Pad90, Katz03, Chav06}. Non-extensive systems  are  spatially inhomogeneous even in equilibrium, their micro-canonical and  canonical ensembles are inequivalent \cite{Thirring, Chav06}, and their entropy function may not be  concave (hence, heat capacities may be negative).
Thin shells  provide a class of analytically solvable models in General Relativity, and they can be employed in order to investigate    axiomatic approaches towards   gravitational thermodynamics \cite{martinez1, martinez}.

In our analysis, we consider static, spherically symmetric thin-shell solutions to Einstein's equations. The thin-shell lies between two regions of Schwarzschild geometry with different masses. The mass $M$ of the external region is the Arnowitt-Deser-Misner (ADM) mass and it can only be positive, while the mass $M_0$ of the interior region can take any real value. For $M_0 > 0$, the interior solution is a black hole; for $M_0 = 0$, the interior solution is flat; and for $M_0 < 0$, the interior solution has a naked singularity.
 
 For a given equation of state, the space of thin-shell solutions is two-dimensional. It is determined by $M$ and $M_0$, or equivalently by $M$ and the temperature at infinity $T_{\infty}$. However, from a thermodynamic point of view, energy and temperature are not independent variables. For example, in the microcanonical distribution, the energy $M$ is fixed and the corresponding temperature is defined as a derivative of the entropy functional. This means that not all solutions correspond to equilibrium configurations. In equilibrium thermodynamics, the latter are  determined by recourse to the Maximum Entropy Principle: an equilibrium configuration is defined as the maximum of the entropy of fixed total energy $M$ \cite{Callen}.

The problem is that in thin shells, as well as  in other self-gravitating systems, matter entropy has no global maxima at fixed energy. This means that the Maximum Entropy Principle does not apply. At face value, this would mean that all thin-shell solutions are thermodynamically unstable: they tend to collapse to their Schwarzschild radius, even in the regime of very small masses. The only way to obtain a consistent equilibrium configuration is by adding an entropy contribution to solutions with negative $M_0$. Because of the simplicity of the system, we can evaluate this entropy term  analytically. It is independent of the shell EoS equation, and
 it is compatible with the corresponding term for solutions to the Tolman-Oppenheimer-Volkoff equation \cite{AnSav12}.

We find that the equilibrium configurations at given $M$ correspond to flat interior solutions ($M_0 = 0$), whenever such solutions exist. When including the Bekenstein-Hawking entropy of horizons for $M_0 > 0$, we find that the system is characterized by four distinct phases, depending on the sign of $M_0$: a regular phase ($M_0 = 0$), a black-hole phase ($M_0 > 0$), and two distinct naked singularity phases ($M_0 < 0$), one characterized by finite temperature at infinity and the other by zero temperature at infinity. We analyze the properties of these phases  and the associated phase transitions.

Finally, we discuss thermodynamic stability in relation to dynamical stability. While solutions of all types can be either dynamically stable or thermodynamically stable, we have found that only solutions with $M_0 = 0$ can be both. In particular, we show that thermodynamically stable solutions with  $M_0 > 0$ fail to be dynamically stable for all EoS. We also have found no solution that is both dynamically and thermodynamically stable for $M_0 < 0$, even though we lack  a  proof that is valid for any EoS.

\medskip

The structure of this paper is the following. In Sec. 2, we analyze the geometry of thin-shell solutions for general equations of state. In Sec. 3, we analyze the thermodynamic consistency of thin-shell solutions,  we identify the entropy term associated to naked singularities, and analyze the thermodynamic phases of the system and their properties. In Sec. 4, we present the thermodynamical states of the system and the analysis of dynamical stability.
In the conclusions, we discuss and summarize our results.

\section{Spacetime geometry for a self-gravitating thin shell}
\subsection{Equilibrium conditions}
A static, spherically symmetric spacetime containing a thin shell of area  $4\pi R^2$ is described by the metric
\begin{equation}
ds^2 = - L(r)^2 dt^2 + \frac{dr^2}{1-\frac{2m(r)}{r}}+r^2(d\theta^2+\sin^2\theta d\phi^2),
\end{equation}
where $L(r)$ is the lapse function, $m(r)$ is the mass function and $(t, r, \theta, \phi)$ is the usual coordinate system.

For $r>R$, the metric is Schwarzschild with ADM mass $M$, i.e.,
\begin{equation}
L(r) = \sqrt{1-\frac{2M}{r}}, \qquad m(r)=M.
\end{equation}
For $r<R$, the the metric is Schwarzschild with  ``mass'' $M_0$, i.e.,
\begin{equation}
L(r) = \frac{\kappa}{\kappa_0}\sqrt{1-\frac{2M_0}{r}}, \qquad m(r)=M_0,
\end{equation}
where
\begin{equation}\label{defk}
\kappa:=\sqrt{1-2M/R},\qquad \kappa_0:=\sqrt{1-2M_0/R}.
\end{equation}
We assume that the ADM mass of the solution is positive, i.e., that $M > 0$. The parameter $M_0$ is defined only in the interior, so it has no interpretation of total energy. It is therefore unconstrained. Hence,  
$\kappa$ takes values in $(0, 1]$, while  $\kappa_0$  takes values in $(0, \infty)$.

 The thin-shell junction conditions \cite{KMM06, Gon02} lead to the structure equations
\begin{align}
&\label{stata} \kappa \, \kappa_0 (1+4P/\sigma) = 1, \\
&\label{statb} \kappa_0-\kappa = 4\pi R\sigma,
\end{align}
where  $\sigma$ and $P$ stand for  the shell's surface  density and pressure,  respectively. They are not independent variables, as they are related by an EoS. We assume that they satisfy the conditions
$\sigma \geq P \geq 0$, and $1 \geq dP/d\sigma \geq 0$.

\subsection{Characterization of solutions} \label{notov}

We use Eqs.  (\ref{defk}) and (\ref{stata}), in order to eliminate the dependence on $\kappa_0$ and $R$ from Eq.  (\ref{statb}). Then, we obtain a fourth-order algebraic equation  for $\kappa$,

\begin{equation}\label{quartic}
\Phi(\kappa):=A\kappa^4 - (1+A)\kappa^2 -B\kappa +1 = 0,
\end{equation}
where $A:=1+4P/\sigma$ and $B:=8\pi M\sigma (1+4P/\sigma)$. In the Appendix \ref{kapp}, we prove that  Eq. (\ref{quartic}) admits a unique solution $\kappa\in (0,1)$ for any $A,B>0$.

Thus the space $Z$ of solutions to Eqs. (\ref{stata}) and (\ref{statb}) is parameterized by the variables $M, A, $ and $B$, i.e.  by the variables $M, \sigma, $ and $P$, the latter two being related by  an EoS.

The elements of $Z$ are of  three types,  depending on the sign of $M_0:=m(0)$.
\begin{itemize}
\item $M_0=0$: Type F (flat): $\kappa_0 = 1$; Minkowski spacetime for $r<R$.
\item $M_0>0$: Type B (black-hole): $\kappa_0< 1$; it contains a Schwarzschild horizon at $r=2M_0$.
\item $M_0<0$: Type S (singularity): $\kappa_0 > 1$; it contains a negative-mass Schwarzschild singularity at $r=0$.
\end{itemize}
F-type solutions are determined by the equation $K(\sigma,P) = M$, where
\begin{equation}\label{kappam}
K(\sigma,P) := \frac{4(1+\frac{2P}{\sigma})}{\pi(1+\frac{4P}{\sigma})^3}\, \frac{P^2}{\sigma^3}.
\end{equation}
Invoking continuity, it is straightforward to prove  that $M<K(\sigma,P)$  for all B-type solutions and   $M>K(\sigma,P)$ for all S-type solutions.

Since $\sigma \geq P$ for all physical equations of state,  $\lim_{P \to \infty} K(\sigma,P) = 0 $. Furthermore, if  $\lim_{P \to 0} P^2/\sigma^3$ is finite, then $K(\sigma,P)$ has  a global maximum $M_{OV}:=\text{max}\,K(\sigma,P)$. For $M > M_{OV}$, all solutions are of type S, i.e., there are no flat solutions. In this sense, $M_{OV}$ is the analogue of the Oppenheimer-Volkoff limit in stellar models, which justifies the notation.

In contrast,  if $\lim_{P \to 0} P^2/\sigma^3=\infty$ then $K(\sigma,P)$ has no global maximum and there exists no Oppenheimer-Volkoff limit: there exist F-type and B-type solutions for all masses $M$.

The asymptotic behavior of the solutions is the following.

\begin{itemize}

\item $\sigma, P \rightarrow\infty$ with constant $P/\sigma$: $B \to \infty$, while $A$ remains constant. Hence, $\Phi(\kappa) \simeq -B \kappa + 1$, so the solution to Eq. (\ref{quartic}) is $\kappa = B^{-1} \sim \sigma^{-1}$.  By Eq. (\ref{stata}), $\kappa_0 = B/A \sim \sigma$, and the solution is of type S.

\item  $\sigma, P \rightarrow 0 $ with constant $P/\sigma$: $\Phi(\kappa) = A\kappa^4 - (1+A)\kappa^2  +1$, which implies that $\kappa = A^{-1/2}$. In this limit, $R \rightarrow 2M\left(1 + \frac{\sigma}{4P}\right)$ is a constant, and $\kappa_0 = A^{-1/2} < 1$, so the solution is of type B.

\item  $\sigma, P \rightarrow 0 $ with $P/\sigma \rightarrow 0$:  $R \rightarrow \infty$, $\kappa=1$ and  $-M \leq M_0 \leq M$ the exact value being determined by the asymptotic behaviour of $\sigma^3/P^2$. For details, see the  Appendix \ref{ntrstng}.

\end{itemize}

\subsection{Stability of solutions}

As shown in Refs.  \cite{Gon02, BLP91}, a thin-shell solution is stable under small perturbations if the function
\begin{equation}\label{defpsi}
\Psi(\kappa,\kappa_0,\beta):=4\kappa^2 \kappa_0^2(1+3 \kappa\, \kappa_0)\,\beta +3\kappa^3 \kappa_0^3 -(\kappa^2+\kappa\, \kappa_0+\kappa_0^2).
\end{equation}
is positive; $\beta = dP/d\sigma$ is the square root of the speed of sound in the shell.
Dynamical stability of thin shell solutions has been studied in \cite{BLP91}, but past studies did not cover the case of $M_0 < 0$. 

\begin{figure}[H]
\centering
\includegraphics[width=13cm]{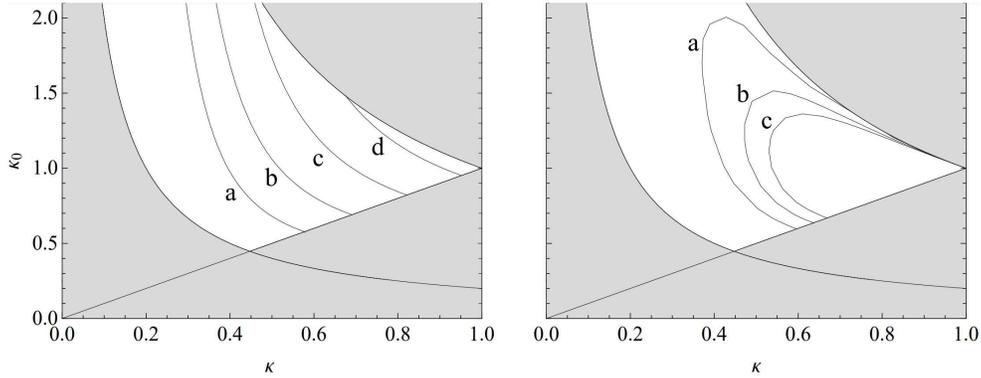}
\caption{ In both plots, the white region describes the portion of the $\kappa-\kappa_0$ plane with physical solutions. This region is defined for $\kappa \in[0, 1]$, $\kappa_0 > 0$, and it is bounded by the line $\kappa = \kappa_0$ and the hyperbolae $\kappa \kappa_0 = 1$ and $\kappa \kappa_0 = 1/5$. Left: The curves $\Psi = 0$ for different constant values of $\beta$; (a) $\beta=1$,  (b) $\beta =1/2$, (c) $\beta =1/5$, and (d) $\beta=1/25$. Right: The curves $\Psi = 0$ for the EoS (\ref{gratton}) for different values of the exponent $\gamma$; (a)$\gamma =1/10$, (b)$\gamma =1/4$ (c) $\gamma=1/3$. }
\label{stabilityplot}
\end{figure}

To analyze dynamical stability, it is convenient to parameterize the space of solutions by $\kappa$ and $\kappa_0$. By definition $\kappa < 1$, and $\kappa < \kappa_0$. Also, by Eq. (\ref{stata}), $\kappa \kappa_0 = A^{-1}$. Since $A = 1 + 4P/\sigma \in [0,5]$, the physical solutions lie between the hyperbolae $\kappa \kappa_0 = 1$ and $\kappa \kappa_0 = 1/5$. The portion of the $\kappa-\kappa_0$ plane with physical solution is drawn in white in Fig. \ref{stabilityplot}, where we also plot the limiting curve defined by Eq. (\ref{defpsi}). Stable solutions lie in the concave side of the limiting curve.
We see that there exist  stable solutions also for $\kappa_0 > 1$.

An interesting feature of the stability analysis is that the F-type solution for $M = M_{OV}$ always lies on the curve $\Psi = 0$.
This is straightforward to show. The maximum of $M$ is obtained by the condition $\frac{d}{d\sigma}K(\sigma, P) = 0$, which implies that $\Psi = 0$.

\subsection{Example: Gratton's equation of state}
Matter in compact stars is expected to be approximately polytropic at sufficiently low temperatures/pressures, while all equations of state are expected to be ultrarelativistic at sufficiently high temperatures/pressures. 
Gratton's EoS \cite{grattoon, gratton}  interpolates between a polytropic EoS at low pressures and 
a linear one at high pressure. 

A large part of our analysis will apply to general equations of state. However, for some calculations or plots, we will work with  the two dimensional analogue of Gratton's EoS, namely,
\begin{equation}\label{gratton}
\sigma = \sigma_r (P/\sigma_r)^\gamma + a P
\end{equation}
where  $\sigma_r$ is some reference density, $a\geq 1$ and $\gamma\in (0, 1)$. For $\sigma\ll \sigma_r$, the EoS
 is polytropic, while for $\sigma\gg \sigma_r$ the linear term prevails.  

In what follows, we use the system of units where  
 $c = G = \sigma_r = k_B = 1$. In these units, all physical magnitudes are dimensionless. We chose this system, rather than the natural one with $\hbar = 1$, because it is more convenient in the, essentially classical, thermodynamic analysis of the system, and because we want to keep $\hbar$ when introducing the Bekenstein-Hawking contribution to entropy, which incorporates quantum effects.

Given an EoS,  the space $Z$ of equilibrium configurations can be parameterized by $(M, P)$. Then, F-type solutions define a curve on the
 $P-M$ plane, {\em the F-curve}.  The F-curve has a global maximum if $\lim_{P\to 0}P^2/\sigma^3$ is finite,  which for EoS (\ref{gratton}) amounts to  $\gamma\leq 2/3$.

\begin{figure}[H]
\centering
\includegraphics[width=12cm]{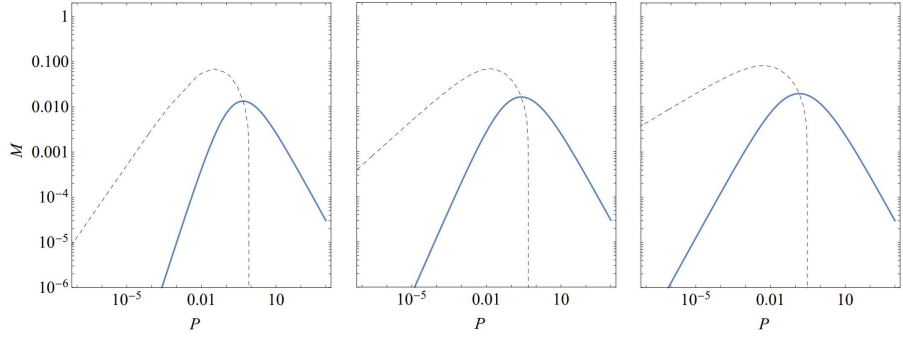}
\caption{The $F$-curve  for different values of  $\gamma < 2/3$, namely $\gamma=1/10$ (left), $\gamma=1/4$ (center), $\gamma=1/3$ (right). B-type configurations lie between the $F$ -curve and the horizontal axis. The remaining configurations are of S-type.  Dynamically stable configurations lie between the dashed curve and the horizontal axis.  The dashed  curve meets the $F$-curve at the maximum.}
\label{fig:fcurve}
\end{figure}

\begin{figure}[H]
\centering
\includegraphics[width=12cm]{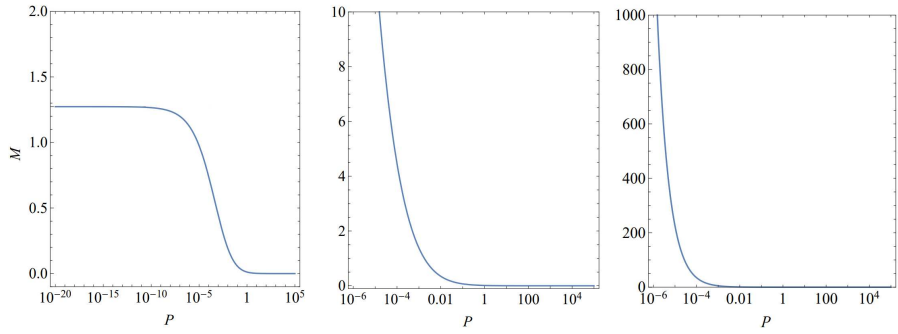}
\caption{The $F$-curve (solid line) for different values of  $\gamma \geq 2/3$, namely $\gamma=2/3$ (left), $\gamma=3/4$ (center), $\gamma=7/8$ (right). B-type configurations lie between the $F$ -curve and the horizontal axis. The remaining configurations are of $S$-type.  There are no dynamically stable configurations. The $F$-curve does not possess  a global maximum with the exception of the marginal case $\gamma=2/3$ where it asymptotically approaches  the limiting value  $M=4/\pi$.}
\label{fig:fcurves}
\end{figure}

In Fig. \ref{fig:fcurve} we plot the $F$-curve for different values of $\gamma < 2/3$, as well as the curve that separates between stable and unstable configurations on the $P-M$ plane. The latter intersects the $F$-curve always at the maximum, i.e. at $M=M_{OV}$. In Fig. \ref{fig:fcurves}, we plot the $F$-curve for different values of $\gamma$ on the $P-M$ plane. In this case there are no dynamically stable configurations. With the exception of the marginal case $\gamma=2/3$ the $F$-curve is unbounded from above. 

\begin{figure}[H]
\centering
\includegraphics[width=6cm]{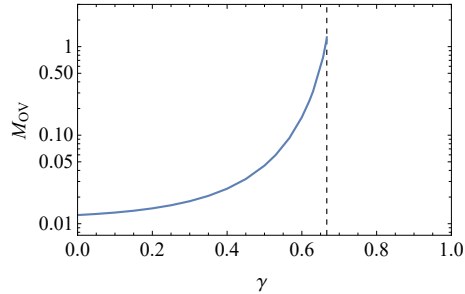}
\caption{The Oppenheimer-Volkoff limit $M_{OV}$ for the Gratton EoS as a function of the parameter $\gamma$.  $M_{OV}$ is not defined  for $\gamma>2/3$.}
\label{fig:MOV}
\end{figure}

The space $V$ is the union of all
one-dimensional manifolds (fibers) $V_{M}$ of constant $M$. There are three types of fibers, which we denote as I, II and III.

\begin{itemize}
    \item Fibers of type I are defined by $\gamma<2/3$ and $M<M_{OV}(\gamma)$.  The line $M =$  constant intersects the $F$-curve
twice. Hence, these fibers involve two F-type solutions, one at $P = P_1$ which is dynamically stable and one at $P = P_2 > P_1$ which is dynamically unstable. Solutions in the interval $(P_1,P_2)$ are B-type. The remaining solutions are of type S.
    \item Fibers of type II are defined by $\gamma<2/3$ and $M>M_{OV}(\gamma)$. They contain only S-type solutions .
    \item Fibers of type III are defined by $\gamma>2/3$.  The line $M =$  constant intersects the $F$-curve
only once, at some point $P_1$. The solutions for $P < P_1$ are B-type solutions, and the solutions for
 $P >P_1$, are S-type. It is straightforward to show that  fiber III  configurations are dynamically unstable.
\end{itemize}

In the following, we will only consider EoS that have no type III fibers. We will therefore employ Gratton's EoS for $\gamma < 2/3$.

\section{Thermodynamic consistency and Entropy of Singularities}
\subsection{Thermodynamic properties of the EoS}
We proceed with an analysis the thermodynamic properties of the EoS.  We assume that the shell consists of a fluid with $k$ particle species. The thermodynamic state space for this fluid is defined by the particle-number densities
 $n_a,\, a=1,2,\dots,k$ and the surface energy density $\sigma$. All thermodynamic properties
  are encoded in the  entropy density function $s(\sigma,n_a)$. The first law of thermodynamics for surface density variables takes the form
\begin{equation}\label{first}
Tds = d\sigma - \sum_a \mu_a dn_a,
\end{equation}
where $T^{-1}:=\big(\frac{\partial s}{\partial \sigma}\big)_{n_a}$ is the local temperature and $\mu_a:=-T\big(\frac{\partial s}{\partial n_a}\big)_{\sigma,\,n_i\neq a}$ is the chemical potential associated to particle species $a$. The pressure $P(\sigma,n_a)$ is then determined via the Euler equation
\begin{equation}\label{euler}
\sigma+P - Ts -\sum_a \mu_a n_a =0.
\end{equation}
Eqs (\ref{first}) and (\ref{euler}) imply the Gibbs-Duhem equation
\begin{equation}\label{gibbs}
dP = s dT + \sum_a n_a d\mu_a = \omega dT + T \sum_a n_a db_a,
\end{equation}
where
\begin{equation}\label{defo}
\omega(\sigma,b_a):= s -\sum_a n_a\bigg(\frac{\partial s}{\partial n_a}\bigg)_{\sigma,\,n_i\neq a}  = s + \sum_a b_a n_a =\frac{\sigma+P}{T},
\end{equation}
and $b_a :=\mu_a/T$ is the {\em activity} of the fluid.

By the maximum entropy principle, the equilibrium configuration for any self-gravitating system is characterized by spatially constant values of the parameters $b_a$ \cite{AnSav21, AnSav14}.
Then, Eqs.   (\ref{gibbs}) and (\ref{defo}) imply that
\begin{equation}\label{dpdt}
\frac{dP}{dT} = \omega = \frac{\sigma+P}{T}  .
\end{equation}
When writing an EoS of state, in which the pressure is a function of the density $\sigma$, it is assumed implicitly that the parameters depend on the activities $b_a$. In particular, this is the case for the parameters $\sigma_r$ and $\gamma$ in Gratton's EoS (\ref{gratton}).

The dominant energy condition $\sigma\geq P$ implies $\frac{\sigma+P}{T} \geq \frac{2P}{T} $. Then Eq (\ref{dpdt}) implies that
\begin{equation}
\frac{d\log P}{d\log T} \geq 2 . \label{const1}
\end{equation}


\subsection{The maximum entropy principle for a thin shell}\label{gravo}
We define the free entropy $\Omega$ of any thermodynamic system as the Legendre transform of the entropy functional $S$ with respect to the total particle numbers for each species $N_a$,  $\Omega = S + \sum_a b_a N_a$. The free entropy is  a  Massieu potential, and a function of $b_a$ and the total energy $M$ of the system. The quantity $\omega$ of Eq. (\ref{defo}) is the (surface) density associated to $\Omega$. The free-entropy representation is natural for self-gravitating systems, because of the constancy of the $b_a$. In this representation, both $P$ and $\sigma$ are functions of  the temperature $T$ and $b_a$, where $T$ varies in the interior of a self-gravitating system. The EoS employed in Einstein's equation is obtained by eliminating $T$. The EoS carries an implicit dependence on the $b_a$'s. For example, the coefficients $\gamma$ and $\sigma_c$ of Gratton's EoS are to be thought as functions of the $b_a$'s.

The free entropy of a spherical thin shell is given by
\begin{equation}\label{oshell}
\Omega_\text{shell}:=4\pi R^2 \omega=4\pi R^2\frac{\sigma+P}{T}.
\end{equation}
The free entropy $\Omega_\text{shell}$ is a function of $M$, $\sigma$ and the activities $b_a$. However, the thermodynamic space for this system depends only on $M$ and $b_a$, $\sigma$ is an unconstrained parameter. In equilibrium thermodynamics, the value of unconstrained parameters is determined by recourse to the Maximum Entropy Principle (MEP). In the  free-entropy representation, the MEP translates into the statement that unconstrained parameters are determined by the maximization of $\Omega$ for fixed mass $M$ and activities $b_a$. Since the $b_a$'s are not affected in the implementation of the MEP, we will drop any reference to them, and consider $\Omega_\text{shell}$ only as a function of $M$ and $\sigma$.

To explain the notion of an unconstrained parameter in this context, note that we can equivalently express $\Omega_\text{shell}$ as a function of $M$ and $M_0$. If we only fix $M$, we assume a self-gravitating system with fixed energy that will move spontaneously into a state of thermal equilibrium. The value of $M_0$ in equilibrium will then be fixed by the MEP. Of course, we expect that the natural equilibrium configuration for sufficiently small mass will be of the F-type, i.e., the interior spacetime will be flat. This will eventually be borne out by our analysis. Analogous properties hold for systems with bulk matter: the space of solutions is larger than the space of equilibrium configurations, the latter being obtained by recourse to the MEP \cite{AnSav12}.

The problem is that $\Omega_\text{shell}$ does not have a global maximum. This is  a generic feature of self-gravitating systems, related to the so-called gravothermal catastrophe. One reason for working with thin shells is that they allow for a simple analytic proof of this statement.

Eq. (\ref{const1}) implies that $P$ diverges at least with $T^2$ for large $T$, and so does $\sigma \geq P$. By Eq. (\ref{quartic}), $\kappa \sim \sigma^{-1}$ in this limit, hence $R$ approaches $2M$. It follows that
$\Omega_\text{shell}$ diverges at least with $T$, for large $T$. Hence, $\Omega_\text{shell}$ has no global maximum and the MEP fails. If this were true, even a thin-shell with very little mass, would spontaneously collapse towards its Schwarzschild radius, in order to maximize entropy.

In Ref. \cite{AnSav12}, it was proposed that analogous problems can be resolved by including  an entropy contribution from the singularities of the S-type solutions. The idea that singularities are entropic objects originates from Penrose. The natural value from the singularity entropy can be inferred from Wald's formulation of black hole entropy in terms of the
Noether charge $Q(\xi)$ of spacetime diffeomorphisms \cite{Wald93},
\begin{eqnarray}
S = \frac{Q(\xi)}{T_{\infty}},
\end{eqnarray}
where $T_{\infty}$ is the temperature measured at infinity. Note that the Noether charge of spacetime diffeomorphisms is always a surface term. The bulk contribution vanishes for any local symmetry, so the Noether charge  is defined on the spacetime boundaries. 

The Noether charge $Q(\xi)$ is defined in terms of the time-like Killing vector $\xi = \frac{\partial}{\partial t}$, normalized so that $\xi^{\mu}\xi_{\mu}=-1$ at infinity, and evaluated on any boundary of the surfaces of constant $t$:
\begin{eqnarray}
Q(\xi) = \frac{\lambda}{4 \pi} \oint_{\partial \Sigma} d\sigma_{\mu \nu} \nabla^{\mu} \xi^{\nu},
\end{eqnarray}
where $\lambda$ is an arbitrary multiplicative constant.

For positive-mass Schwarzschild spacetime, $Q(\xi) = 2 \lambda M$, when evaluated at the horizon. Since $T_{\infty} = 1/(8 \pi M)$, the Bekenstein-Hawking entropy $S_{BH} = 4\pi M^2$ is obtained for $\lambda = \frac{1}{4}$. For negative-mass Schwarzschild spacetime, the singularity at $r = 0$ defines a timelike boundary, for which $Q(\xi) =  2 \lambda M_0\kappa/\kappa_0$. There is no first-principles derivation of the constant $\lambda$. However, the analysis of different systems has shown that the value $\lambda = 2$ is the only one that allows for a consistent implementation of the MEP in self-gravitating systems. Notably, this value works for different EoS, and it is not model dependent.

Here, for $\kappa_0 > 1$, we add a term
\begin{equation}\label{defosing}
\Omega_\text{sing}:= \frac{2 \lambda M_0}{T \kappa_0}=\frac{ \lambda R}{T}\bigg(\frac{1}{\kappa_0}-\kappa_0\bigg),
\end{equation}
to the free energy of the shell, so that the total free entropy $\Omega$ of the system is
\begin{equation}\label{ototplusosing}
\Omega:= \Omega_\text{shell} + \theta(\kappa_0 -1) \Omega_\text{sing} = \frac{R}{T}\big[4\pi R(P+\sigma)+ \lambda \theta(\kappa_0 -1) (\kappa_0^{-1}-\kappa_0)\big],
\end{equation}
where $\theta(x)$ is the step function. There is zero contribution from $\Omega_\text{sing}$ for $\kappa_0 \leq 1$.

Hence, by Eq. (\ref{statb})
$$
\Omega = \frac{R}{T}\big[4\pi R(P+\sigma)+ \lambda \theta(\kappa_0 -1) (\kappa_0^{-1}-\kappa-4\pi R\sigma)\big].
$$
For large $T$,  $R\to 2M$ and  $(\frac{\lambda}{\kappa_0}-\lambda\kappa) \sim \sigma^{-1}\to 0 $.  Hence,  $\Omega$ behaves asymptotically as
\begin{equation}\label{OlrgT}
\Omega_{\infty}= 16\pi M^2\frac{(1-\lambda)\sigma+P}{T}.
\end{equation}
By the dominant energy condition $P \leq \sigma$, hence, $\Omega_{\infty} \leq 16\pi M^2\frac{(2-\lambda)\sigma}{T}$.  For $\lambda > 2$, $\Omega_{\infty} \rightarrow - \infty$, and the MEP is restored. However, the fact that $\Omega$ becomes unbounded from below is  unphysical, as it would imply negative entropy, in contradiction to the statistical interpretation of entropy\footnote{A negative value of entropy that is bounded from below would not be a problem, as the MEP is invariant under an affine transformation $S \rightarrow a S + b$, for $a> 0$, that can be used to always render the entropy positive.}. The case $\lambda = 2$ is the only one possible that can lead to a finite value of $\Omega_{\infty}$, and be compatible with both the MEP and the positivity of entropy. Hence, as in the case of bulk gravitating system, the value of $\lambda = 2$ for the entropy of the singularity is preferred.

We must note, however, that the case for $\lambda = 2$ in thin shells is slightly weaker than the corresponding case of bulk matter\footnote{The reason for this difference is that in general the EoS for a two dimensional system cannot be read from the EoS from a three-dimensional system at a limit where one dimension vanishes. Hence, results obtained with a thin shell of finite width with a 3-d EoS do not coincide with the results obtained from a genuinely 2-d shell.}. Suppose that the  asymptotic ratio $P/\sigma$ in the EoS is a constant, which we denote by $a$. The dominant energy condition implies that $a \leq 1$, while by the positivity of pressure, $a > 0$. Then, asymptotically,
$\Omega_{\infty} = 16\pi M^2\frac{(1 + a -\lambda)\sigma}{T}$. It follows that the preferred value of $\lambda$ is $1 + a$. We have to assume a universal asymptotic ratio $a = 1$, in order to derive $\lambda = 2$. While the existence of  such a ratio appears plausible, we have found no fundamental justification. In practice, this is not a constraint. If an EoS applies up for all temperatures up  to a $T_1$, we can simply extrapolate it for $T > T_1$, so that the asymptotic behavior $P = \sigma$ applies in the limit $T \rightarrow \infty$. Then, by construction, $\Omega_{\infty} = 0$ for $\lambda = 2$.

It is straightforward to show that $\Omega$ is either finite or it vanishes at the limit $T \rightarrow 0$. Hence, there is no problem with the implementation of  the MEP at low temperatures.

\subsection{F-type solutions}
For F-type solutions,  $\kappa = A^{-1}$ where $A = 1 + 4P/\sigma$. By Eq. (\ref{quartic}), $B = (A-1)^2(A+1)/A^2$, and consequently $8 \pi M \sigma =   (A-1)^2(A+1)/A^3$.
We straightforwardly calculate
\bey
\Omega_{shell} = \frac{16\pi M^2  \sigma (1 + P/\sigma)}{T} = \frac{M}{T} \frac{A(A+3)}{2 (A+1)}.
\eey

Since $A\in [1, 5]$, we obtain the bounds
\bey
\frac{M}{T} \leq \Omega_{shell} \leq \frac{10}{3} \frac{M}{T}. \label{ineqO}
\eey
Analogous bounds exist for general self-gravitating regular solutions to Einstein's equations.

A key result of our analysis is that the maxima of $\Omega$ for fixed $M$ correspond to type-F solutions. To show this, we note that for F-type solutions,
\bey
\delta \Omega_{shell} &=& \Omega_{shell}\left( \frac{4A}{A^2-1} \delta \kappa + \frac{\delta \sigma}{\sigma + P}\right), \label{dshell}\\
\delta \Omega_{sing} &=& -2 \lambda \frac{R}{T} \theta (\delta \kappa_0) \delta \kappa_0.
\eey
We proceed to express $\delta \Omega_{shell}$ in terms of $\delta \kappa_0$. To this end, we note that $\delta \kappa$ for  a static configuration is obtained by the requirement that $\frac{\partial \Phi}{\partial \kappa} \delta \kappa + \frac{\partial \Phi}{\partial A} \delta A + \frac{\partial \Phi}{\partial B} \delta B = 0$. From the definition of $B$, $\delta B = (B/\sigma)\delta \sigma + (B/A) \delta A$. Also for F-type solutions $\kappa=A^{-1}$ hence Eq (\ref{quartic}) implies $B=\frac{(A-1)^2(1+A)}{A^2} $. We thus obtain
\bey
\delta \kappa = -\frac{(1+A)}{A(3+2A+A^2)}\delta A - \frac{(A-1)(A+1)}{A(3+2A+A^2)} \frac{\delta \sigma}{\sigma}. \label{dkappa}
\eey
We can also express $\delta\sigma/\sigma$ in terms of $\delta A$,
\bey
\frac{\delta \sigma}{\sigma} =  \frac{\delta A}{4\beta + 1 - A}, \label{dsigma}
\eey
where $\beta  = \frac{dP}{d \sigma}$. By Eq. (\ref{stata}), $\delta \kappa_0 = - A \delta \kappa - \frac{\delta A}{A}$. Using Eqs. (\ref{dkappa}) and (\ref{dsigma}), we obtain
\bey
\delta \kappa_0 = - \frac{A^3+A^2+A-3-12\beta - 4\beta A}{A(A-1-4 \beta)(3+2A+A^2)} \delta A. \label{dkappa0}
\eey
We substitute Eq. (\ref{dkappa0}) and (\ref{dsigma}) into Eq. (\ref{dshell}),  to obtain the remarkably simple result,
\bey
\delta \Omega_{shell} = \Omega_{shell} \frac{4A}{(A+3)(A-1)} \delta \kappa_0 = \frac{R}{T} \delta \kappa_0.
\eey
Hence, for any F-type solution,
\bey
\delta \Omega = \frac{R}{T} [1 - 2\lambda \theta(\delta \kappa_0) ]\delta \kappa_0.
\eey
It follows that  $\delta \Omega < 0$, for all $\delta \kappa_0$, as long as $\lambda > \frac{1}{2}$.
The introduction of the singularity entropy   guarantees that type F solutions are local maxima of $\Omega$.

\subsection{Example: Gratton EoS}
Next, we demonstrate our results above for  the EoS (\ref{gratton}). By Eq. (\ref{dpdt}),
\bey
P=\big[\frac{(T/T_c)^{(1+a)(1-\gamma)}-1}{1+a}\big]^{\frac{1}{1-\gamma}}.
\eey
where $T_c$ is a constant of integration. The EoS cannot be extended to temperatures smaller than $T_c$. In principle, this means that the EoS (\ref{gratton}) fails as $T$ approaches $T_c$, and we have to deform it so that the functions $P(T)$ and $\sigma(T)$ can be extended to $T = 0$. Since $T_c$ is an arbitrary constant, we can take it  so small that the EoS applies arbitrarily close to $T = 0$. Then, the free entropy becomes
\begin{equation}
\Omega= 16\pi M^2\bigg[\frac{(1+a-a \lambda)T}{(1+a)^\frac{1}{1-\gamma}T_c^{1+a}}+\frac{(1-\lambda)T^{\gamma(1+a)-1}}{(1+a)^\frac{\gamma}{1-\gamma}T_c^{\gamma(1+a)}} \bigg].
\end{equation}
The asymptotic behaviour of $\Omega$ at large $T$ for $\lambda = 2$ is shown in  Table \ref{table:1}. The MEP fails only for $a < 1$, in which case the EoS violates the dominant energy conditions. The case $a > 1$ violates entropy positivity, since $\Omega$ is unbounded from below. The only physical case corresponds to the asymptotic behavior $P = \sigma$ for the EoS. However, even in this case, subdominant terms in $\Omega$ lead to violation of entropy positivity, unless $\gamma \leq \frac{1}{2}$.

\begin{table}
\begin{center}
\begin{tabular}{ c| c c c}
   & $\gamma<\frac{1}{ 2}$ & $\gamma =\frac{1}{2}$  & $\gamma>\frac{1}{2}$  \\ [1ex]
\hline   \\ [-2ex]
$a<1$ & $+\infty$ & $+\infty$& $+\infty$ \\  [1ex]
$a=1$ & $0$ & $-8\pi M^2 / T_c^2$& $-\infty$ \\  [1ex]
$a>1$&$-\infty$ &$-\infty$ & $-\infty$
\end{tabular}
\end{center}
\caption{Asymptotic behaviour of $\Omega$ for large $T$ and  $\lambda=2$.}
\label{table:1}
\end{table}

\begin{figure}[H]
\centering
\includegraphics[width=10cm]{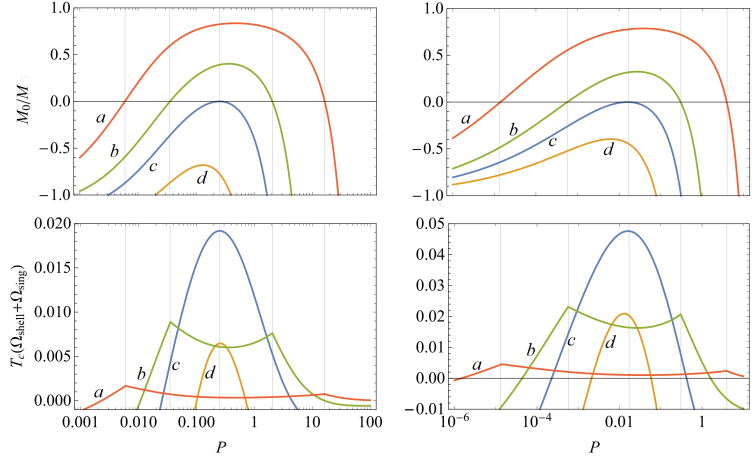}
\caption{Flat configurations, when available, coincide with the local maxima
of $\Omega_\text{shell} +\Omega_\text{sing}$.  Here, we plot $M_0/M$ and $(\Omega_\text{shell} +\Omega_\text{sing})T_c$ against $P$ for different values of $M$ and $\gamma$.
 Left: $\gamma=1/4$ and (a) $M=0.0016$, (b) $M=0.008$, (c) $M =M_{OV} \simeq 0.0162$, (d) $M=0.032$ Right: $\gamma=1/2$ and (a) $M=0.0045$, (b) $M=0.0227$, (c) $M =M_{OV} \simeq 0.0453$, (d) $M=0.0907$.}
\label{fig:regmax}
\end{figure}

As  shown in Sec. 3.3, F-type solutions correspond to local  maxima of the free entropy $\Omega$, for all EoS. For Gratton's EoS, this  behavior is demonstrated graphically in Fig. \ref{fig:regmax}. However, not all local maxima are F-type solutions. As explained in Sec. 2.3, there exist no F-type solutions for $M > M_{OV}$, hence, for $M > M_{OV}$ local maxima are S-type solutions.

\begin{figure}[H]
\centering
\includegraphics[width=8cm]{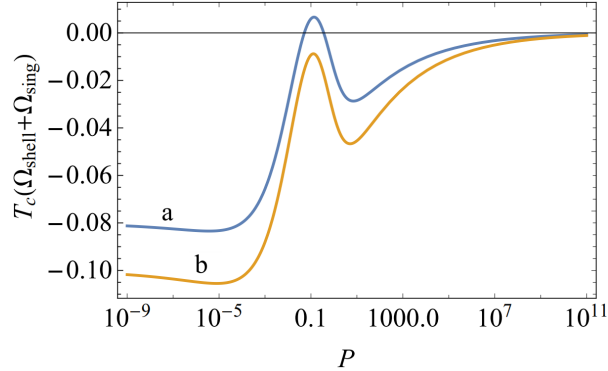}
\caption{
Beyond the TOV limit,  $\Omega_\text{shell}+\Omega_\text{sing}$ is maximized by configurations of  type either  $S_{I}$ or $S_{II}$. Here we plot   $T_c (\Omega_\text{shell}+\Omega_\text{sing})$ against $P$ for $\gamma=1/3$ and  (a) $M=0.04$, (b) $M=0.05$ to display a transition between SI and SII solutions as global entropy maxima.}
\label{fig:pt}
\end{figure}

By Eq. (\ref{ineqO}), F-type solutions have strictly positive free entropy $\Omega$. This is not the case for S-type solutions, even the equilibrium configurations may have $\Omega < 0$. In this case, the global maximum of entropy is $\Omega_{\infty} = 0$, and it is achieved for $T \rightarrow \infty$---see Fig. \ref{fig:pt}. Strictly speaking, this means that the MEP fails in this situation. However, the entropy difference between such solutions is negligible, and they all correspond to the same geometry: the shell arbitrarily close to the horizon, very large energy density balanced by a large negative value of $M_0$. Even if the local temperature $T$ diverges, the entropy at infinity $T_{\infty} = T \kappa$ tends to zero.
If we assume  a length cut-off in the admissible proper distance of the shell from its Schwarzschild  horizon (say the Planck length), then there is indeed a state of maximal entropy very close to zero, that describes a thin shell at almost Schwarzschild radius. They are broadly similar to the configurations proposed for the membrane paradigm or the brick wall model for black holes, the main difference being that they are objects of almost zero entropy and temperature.
 We will refer to such asymptotic equilibrium solutions
as being of the SII type; we will refer to S-type solutions at finite temperature as SI solutions.

\section{Thermodynamic phases and stability}
\subsection{Phase diagrams}
In the previous section, we showed that the introduction of a singularity entropy term is necessary in order to (i) enable the implementation of the MEP, and (ii) to guarantee that F-type solutions are equilibrium configurations for low ADM masses, in accordance with our expectations from Newtonian gravity. However, we ignored the entropy contributions from the horizon of the B-type solutions, i.e., their Bekenstein-Hawking entropy. Without an entropy contribution from the horizon, B-type solutions cannot dominate thermodynamically, as seen in Sec. 3, only F-type and S-type solutions maximize entropy.

We incorporate  the Bekenstein-Hawking entropy, by adding a term  
\begin{equation}\label{otota}
\Omega_{\text{BH}}:=4\pi M_0^2 /\hbar
\end{equation}
to the free energy for $M_0 > 0$. This is the Bekenstein-Hawking entropy associated to the horizon inside the black hole. 
The total free energy of a solution takes the form
\begin{equation}\label{otota}
\Omega =
\begin{cases}
&\Omega_{\text{matt}}+\Omega_{\text{BH}}\,,\quad M_0>0 \\
&\Omega_{\text{matt}} + \Omega_{\text{sing}}\,,\quad M_0<0.
\end{cases}
\end{equation}

With this expression for free energy, the free entropy along a fiber of constant $M$ behaves as follows.

\begin{itemize}
\item For $M < M_{OV}$, there are four local maxima of $\Omega$, two correspond to F-type solutions, one to a B-type solution and one to an SII solution.
\item For $M > M_{OV}$, there are two local maxima of $\Omega$, one at a SI solution and one at a SII solution.
\end{itemize}

In equilibrium thermodynamics, local maxima of entropy correspond to distinct phases, the {\em global maximum} defining the equilibrium phase. Keeping this in mind, we  identify the thermodynamic phases for a shell described by the Gratton EoS (\ref{gratton}). The thermodynamic state space is defined by $M$, but also by the dimensionless parameters $\gamma$ and $\nu := \sigma_c^2 G^3  c^{-11} \hbar$, which are (in general) functions of the activities.

\begin{figure}[H]
\centering
\includegraphics[width=12cm]{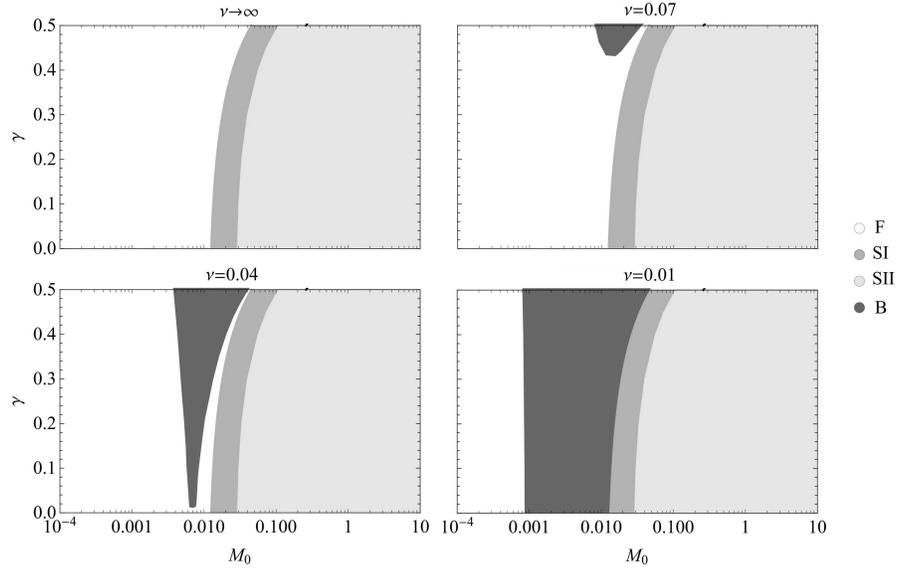}
\caption{Phase diagram for a thin shell with matter described by the EoS (\ref{gratton}) for $a = 1$. The thermodynamic state space depends on the ADM mass, the exponent $\gamma$ and the dimensionless parameter
$\nu := \sigma_c^2 G^3  c^{-11} \hbar$.}
\label{fig:bhphd}
\end{figure}

The phase diagram is given in Fig. (\ref{fig:bhphd}). For small M, the equilibrium phase is always of F type, in agreement with our intuitions from Newtonian gravity. The B phase vanishes for $\nu \rightarrow \infty$, and it becomes increasingly dominant as $\nu \rightarrow 0$. For $M> M_{OV}$, there are only S-type solutions: as long as the free energy remains positive, the equilibrium phase is SI, for larger values of $M$ it is SII. Note that there is no coexistence curve between the B and the S phase: since the B phase corresponds to $M_0>0$ and the S phase to $M_0 < 0$, we always encounter an F phase at $M_0 = 0 $ when going from B to S. For large $\nu$, the interpolating F phase is so small that it cannot be distinguished.

\begin{figure}[H]
\centering
\includegraphics[width=10cm]{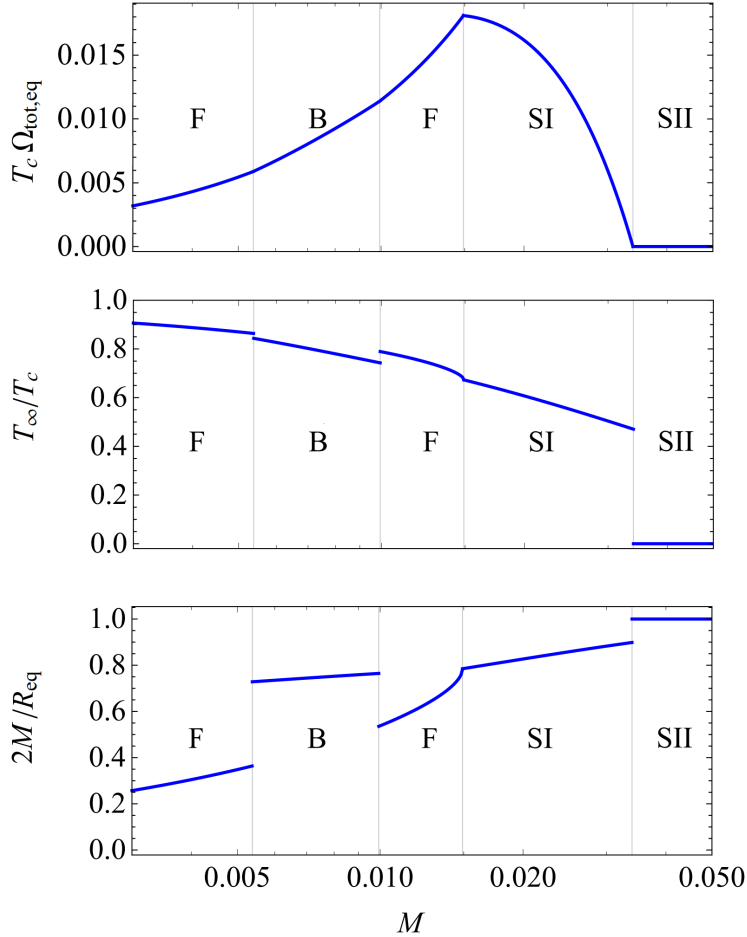}
\caption{ The free entropy $\Omega$,  temperature $T_{\infty}$ at infinity and the shell radius $R$ for equilibrium solutions as a function of the ADM mass $M$ for  $\gamma=0.2$ and $\nu=0.04$. }  
\label{fig:tinfb}
\end{figure}

In Fig.  \ref{fig:tinfb}, we plot the temperature at infinity $T_{\infty}$ and the free entropy $\Omega_{eq}$ for the equilibrium solutions as a function of the ADM mass $M$. The free-entropy is, by construction, a continuous function of M, with discontinuous derivatives at the transitions. The temperature at infinity may be discontinuous at the transition points, the discontinuity $\Delta T_{\infty}$ defining the latent heat of the transition $\ell = \Omega_{eq}\Delta T_{\infty}$. Fig. \ref{fig:tinfb} demonstrates that $\ell\neq 0$ at the B-F and the SI-SII boundaries, suggesting that the corresponding transitions are first-order; and that $\ell = 0$ at the F-SI boundary, suggesting a continuous phase transition.

The discontinuity of the temperature at infinity in Fig.  \ref{fig:tinfb} owes to the fact that transitions $F\to B$, $B\to F$ and $S_I\to S_{II}$ are first-order. In these transitions, $\Omega(M,P)$ has two local maxima for constant $M$, one of which is a global maximum. The specification of the global maximum changes at the transition point. This behavior 
 is depicted in Fig.5 for  $S_I\to S_{II}$ transitions.

We also note that $T_{\infty}$ is a decreasing function of mass $M$, hence, the associated heat capacity $C = \frac{dM}{dT_{\infty}}$ is negative. This has no impact on thermodynamic stability. First, we note that the heat capacity associated to stability is defined as a derivative at constant particle numbers, while $C$ defined here is defined at constant activities. Second, thermodynamic stability is guaranteed by the concavity of the equilibrium free entropy  $\Omega_{eq}$, which follows from the use of the MEP in the latter's construction.

\subsection{Dynamical stability}

To analyze  dynamical stability, we evaluate the function $\Psi$ of Eq. (\ref{defpsi}) for solutions that maximize entropy maxima. We can prove that for any EoS, the equilibrium solutions of type B lie on the curve $\Psi = 0$. The proof is provided in the Appendix \ref{psi0}. A solution with $\Psi = 0$ is dynamically unstable, because small perturbations around equilibrium are guided by a quadratic term and they are unbounded in one direction. Hence, there are no B-type solutions that are both dynamically and thermodynamically stable. In contrast, F-type equilibrium solutions are also dynamically stable, with $\Psi > 0$.

\begin{figure}[H]
\centering
\includegraphics[width=9cm]{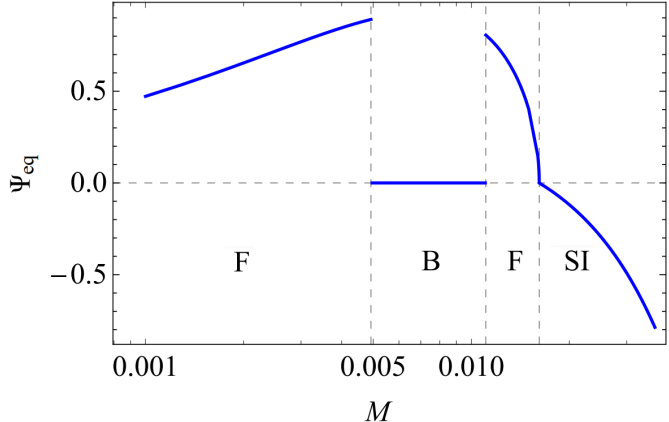}
\caption{The  function $\Psi$  for equilibrium solutions with the Gratton EoS for $\gamma=\frac{1}{4}$ and $\nu= \frac{1}{25}$.  Only F-type solutions are both dynamically and thermodynamically stable. }
\label{psiplot}
\end{figure}

As mentioned earlier, S-type solutions maximize the total free energy only for $M > M_{OV}$. Solutions of type SII are always unstable, because $\Psi < 0$ as $\kappa_0 \rightarrow \infty$. We have not been able to provide a proof valid for a general EoS for the sign of $\Psi$ in solutions of type SI.
However, we have found $\Psi < 0$ for all S-type solutions that maximize entropy with the Gratton EoS. A representative plot is given in Fig. \ref{psiplot}.
We conjecture that the S-phase is generically dynamically unstable, even though we cannot preclude the existence of small regions of stability for the SI phase, especially near $M = M_{OV}$.

If this conjecture is true, then only F-type solutions can be {\em both} dynamically and thermodynamically stable. Again this is another result, where the behavior of thin-shell solutions differs from the behavior of self-gravitating solutions with bulk matter.

\section{Conclusions}
In this article, we gave a general analysis of the thermodynamics of spherically symmetric thin-shell solutions to Einstein's equations. We included solutions with negative interior mass, which have not been analyzed in the literature. We showed that the inclusion of such solutions is essential for the thermodynamic consistency of the system. We can only implement the Maximum Energy Principle when we include an entropy term from the singularity that corresponds to these solutions, in addition to the Bekenstein-Hawking term for the entropy of solutions with positive interior mass. Then, we can formulate a consistent thermodynamic description for this system, and identify,  four distinct thermodynamic phases. Our results are fully consistent with analogous results obtained for bulk thermodynamic systems. In particular, they provide an additional argument in support of the assignment of entropy to the singularity of negative-mass Schwarzschild spacetime, and, consequently, to Penrose's conjecture about the assignment of entropy to singularities.

Thin shell solutions provide paradigmatic systems by which to explore the properties of gravitational equilibrium thermodynamics. As shown in this paper, most thermodynamic properties of such systems can be demonstrated analytically. The next step will be to employ non-static thin shell models in order to analyze non-equilibrium thermodynamic processes, in particular, the manifestation of a generalized second law of thermodynamics in relation to black hole and naked singularity solutions analyzed here.

\section*{Acknowledgements}
D.K. acknowledges financial support from the “Andreas Mentzelopoulos Foundation”.

\appendix

\section{Uniqueness of solutions to Eq.   (\ref{quartic})} \label{kapp}

We prove that Eq. (\ref{quartic}) admits a unique solution $\kappa\in (0,1)$ for any $A,B>0$.

  By definition, $\Phi'(\kappa) = 4A\kappa^3 - 2(1+A)\kappa -B$. If  $\Phi'(1) \leq 0$, then $4A\leq  2(1+A)+B$. Hence, $4A\kappa^3\leq  2(1+A)\kappa^3+B\kappa^3<2(1+A)\kappa+B$, for all $\kappa\in(0,1)$. It follows that  $\Phi' < 0$ in $ (0,1)$, and $\Phi$ possesses at most one root for $\kappa \in (0,1)$.

If  $\Phi'(1) > 0$, then $\Phi'(0)\Phi'(1) < 0$ since $\Phi'(0) = -B<0$.  There exists  $\kappa_*\in(0,1)$ such that $\Phi'(\kappa_*) = 0$, and  $\Phi'(\kappa) > 0$ for all $\kappa\in(\kappa_*,1)$, so that $\Phi$ possesses {\em no root}  in $[\kappa_*,1)$. By definition $4A\kappa_*^3 = 2(1+A)\kappa_* + B$, hence, $4A\kappa^3 = 2(1+A)\kappa\frac{\kappa^2}{\kappa_*^2} + B\frac{\kappa^3}{\kappa_*^3} < 2(1+A)\kappa + B $,  for all $\kappa\in(0,\kappa_*)$. Thus $\Phi' <0$  for any in $(0,\kappa_*)$. Hence $\Phi(\kappa)$ possesses at most one root  in $ (0,\kappa_*)$, and consequently, in $(0,1)$.

Since  $\Phi(0)=1>0$ and $\Phi(1) =- B<0$, $\Phi$ possesses at least  one root in $(0,1)$. We conclude that $\Phi$ possesses {\em a single root} in $ (0,1)$.

\section{Asymptotic behaviour in the $\sigma, P \rightarrow 0 $ with $P/\sigma \rightarrow 0$ limit} \label{ntrstng}

Eq (\ref{quartic}) can be cast as
\begin{equation}\label{appqa}
A\varepsilon^4 -4 A\varepsilon^3+(5A-1)\varepsilon^2 + (2-2A+B)\varepsilon -B=0,
\end{equation}
where $\varepsilon:=1-\kappa>0$. If $\sigma, P \rightarrow 0 $ with $P/\sigma \rightarrow 0$ then Eq (\ref{appqa}) cannot hold unless   $ \varepsilon \ll 1$, whence it  becomes
\begin{equation}
(5A-1)\varepsilon^2 + (2-2A+B)\varepsilon -B=0.
\end{equation}
Since $\varepsilon > 0$ the latter implies
\begin{equation}\label{appqb}
8\varepsilon =2(A-1)-B+\sqrt{16B+B^2+16B(A-1)+4(A-1)^2}
\end{equation}
Since $B\ll 1$ the $-B$ term on the r.h.s. of Eq (\ref{appqb}) is negligible with respect  to the square root which falls at most with $\sqrt{B}$. Hence Eq (\ref{appqb}) becomes
\begin{equation}\label{appqc}
8\varepsilon =2(A-1)+\sqrt{16B+B^2+16B(A-1)+4(A-1)^2}.
\end{equation}
Since $|1-A|, B\ll 1$ terms $B^2$ and $16B(A-1)$ are negligible with respect to $16B$ and Eq (\ref{appqc})  becomes
\begin{equation}\label{appqd}
8\varepsilon =2(A-1)+\sqrt{16B+4(A-1)^2}
\end{equation}
Since $A:=1+4P/\sigma$ and $B:=8\pi M \sigma(1+4P/\sigma)$ Eq (\ref{appqd}) becomes
\begin{equation}\label{appqe}
\frac{\varepsilon}{4P/\sigma} = \frac{1}{4}+\frac{1}{4}\sqrt{1+8\pi M\frac{\sigma^3}{P^2}\bigg(1+\frac{4P}{\sigma}\bigg)}.
\end{equation}

We see that
\begin{equation}
\frac{4P/\sigma}{\varepsilon} =
\begin{cases}
&2,\,\text{if}\, \lim_{\sigma\to 0}\sigma^3/P^2 =0 \\
&p,\,\text{if}\, \lim_{\sigma\to 0}\sigma^3/P^2 \neq 0,\infty \\
&0,\,\text{if}\, \lim_{\sigma\to 0}\sigma^3/P^2 =\infty,
\end{cases}
\end{equation}
where $0<p<2$ some finite constant.

On the other hand Eq(\ref{defk}) implies
\begin{equation}\label{mtom}
\frac{M_0}{M} = \frac{1-\kappa_0^2}{1-\kappa^2}  = \frac{(1+\kappa_0)(1-\kappa_0)}{(1+\kappa)(1-\kappa)}=\frac{(1+\frac{1}{A\kappa})(1-\frac{1}{A\kappa})}{(1+\kappa)(1-\kappa)}.
\end{equation}
In the limit under consideration $\frac{1+\frac{1}{A\kappa}}{1+\kappa}=1$. Also by definition $
\frac{1-\frac{1}{A\kappa}}{ 1-\kappa} =  \frac{ 1-\frac{1}{(1+4P/\sigma)(1-\varepsilon)}}{ \varepsilon}$ hence in the same limit $\frac{1-\frac{1}{A\kappa}}{ 1-\kappa} = \frac{4P/\sigma}{\varepsilon}-1$. Thus Eq (\ref{mtom}) implies
\begin{equation}\label{mtomb}
\frac{M_0}{M} = \frac{4P/\sigma}{\varepsilon}-1 =
\begin{cases}
&1,\,\text{if}\, \lim_{\sigma\to 0}\sigma^3/P^2 =0 \\
&p-1,\,\text{if}\, \lim_{\sigma\to 0}\sigma^3/P^2 \neq 0,\infty \\
&-1,\,\text{if}\, \lim_{\sigma\to 0}\sigma^3/P^2 =\infty.
\end{cases}
\end{equation}

\section{$\Psi$ vanishes for equilibrium B-type solutions} \label{psi0}

The differentiation of  Eq. (\ref{quartic}) with respect to $\sigma$ for generic EoS $P=P(\sigma)$ implies
\begin{equation}
\frac{\partial \kappa}{\partial \sigma}\bigg|_M =
\frac{-2 M \pi \sigma^2 \kappa - 8 M \pi \sigma^2 \beta \kappa  - \sigma \beta \kappa^2 +
 \sigma \beta \kappa^4 + \kappa^2 P -
 \kappa^4 P}{\sigma (2 M \pi\sigma^2 + \sigma\kappa -\sigma \kappa^3+
   8 M\pi\sigma P + 2\kappa P - 4\kappa^3P)}
\end{equation}
Eqs (\ref{stata}) and (\ref{statb}) can be used to eliminate $\sigma,P$ thus yelding
\begin{equation}\label{dkdx}
\frac{\partial \kappa}{\partial \sigma}\bigg|_M = \frac{8 \kappa^2 M \pi [ \kappa_0^2 (1 + 4 \beta)-1 ]}{(\kappa - \kappa_0) (\kappa - 3 \kappa^3 + \kappa_0 +
   \kappa^2 \kappa_0)}
\end{equation}
Similarly, we obtain
\begin{equation}\label{dkodx}
\frac{\partial \kappa_0}{\partial \sigma}\bigg|_M =
-\frac{8 \pi \kappa_0 M  (-2 \kappa^3 + \kappa_0 + 3 \kappa^4 \kappa_0 - 2 \kappa^3 \kappa_0^2 - 4 \kappa^2 \kappa_0 v +
    12 \kappa^4 \kappa_0 v - 8 \kappa^3 \kappa_0^2 v)}{(\kappa-1 ) (1 + \kappa) (\kappa - \kappa_0) (-\kappa +
    3 \kappa^3 - \kappa_0 - \kappa^2 \kappa_0)}.
\end{equation}
It is then straightforward to show that $\frac{\partial \Omega_\text{bh}}{\partial \sigma}\big|_M=0$ if $\frac{\partial}{\partial \sigma}\frac{M_0}{M}\big|_M=0$, i.e. if
\begin{equation}\label{ddx}
\frac{\partial}{\partial \sigma} \frac{1-\kappa_0^2}{1-\kappa^2} \bigg|_M=0.
\end{equation}
We use Eqs. (\ref{dkdx}) and (\ref{dkodx}) in order to eliminate $\frac{\partial\kappa}{\partial\sigma}\big|_M$ and  $\frac{\partial\kappa_0}{\partial\sigma}\big|_M$ on the l.h.s. of Eq. (\ref{ddx}). Then, we obtain
\begin{equation}\label{zeropsi}
\beta =\frac{\kappa^2+\kappa_0\kappa+\kappa_0^2-3\kappa_0^3\kappa^3}{4\kappa_0^2\kappa^2(1+3\kappa_0\kappa)}.
\end{equation}
It follows that $\frac{\partial \Omega_\text{bh}}{\partial \sigma}\big|_M=0$ when $\Psi(\kappa,\kappa_0,\beta)=0$.\\

By Eq. (\ref{oshell}),
\begin{equation}
\frac{1}{\Omega_\text{shell}}\frac{\partial \Omega_\text{shell}}{\partial \sigma}\bigg|_M = \frac{2}{R}\frac{\partial R}{\partial \sigma}\bigg|_M +  \frac{1}{\omega}\frac{\partial \omega}{\partial \sigma}\bigg|_M.
\end{equation}
Since $R=\frac{2M}{1-\kappa^2}$
\begin{equation}
 \frac{2}{R}\frac{\partial R}{\partial \sigma}\bigg|_M   =  \frac{4\kappa}{1-\kappa^2}  \frac{\partial \kappa}{\partial \sigma}\bigg|_M=\frac{32 \pi M \kappa^3  (-1 + \kappa_0^2 + 4 \kappa_0^2 \beta)}{(-1 + \kappa) (1 + \kappa) (\kappa -
   \kappa_0) (-\kappa + 3 \kappa^3 - \kappa_0 - \kappa^2 \kappa_0)},
\end{equation}
where in the last step we used Eq. (\ref{dkdx}) in order to eliminate $ \frac{\partial \kappa}{\partial \sigma}\big|_M$.

We also calculate
\begin{equation}
\frac{1}{\omega}\frac{\partial \omega}{\partial \sigma}\bigg|_M = \frac{1}{\sigma+P} = \frac{32\pi M \kappa \kappa_0}{(\kappa^2-1)(\kappa-\kappa_0)(1+3\kappa \kappa_0)},
\end{equation}
where we eliminated $\sigma$ and $P$ using  Eqs. (\ref{stata}) and (\ref{statb}). Then,
\begin{equation}
\frac{1}{\Omega_\text{shell}}\frac{\partial \Omega_\text{shell}}{\partial \sigma}\bigg|_M = \frac{32 \pi  M \kappa  (-\kappa ^2 - \kappa  \kappa _0 - \kappa _0^2 + 3 \kappa ^3 \kappa _0^3 + 4 \kappa ^2 \kappa _0^2 \beta +
   12 \kappa ^3 \kappa _0^3 \beta)}{(-1 + \kappa ) (1 + \kappa ) (\kappa  - \kappa _0) (1 + 3 \kappa  \kappa _0) (-\kappa  +
   3 \kappa ^3 - \kappa _0 - \kappa ^2 \kappa _0)}
\end{equation}
We see that  $\frac{\partial \Omega_\text{shell}}{\partial \sigma}\big|_M = 0$ when $\Psi(\kappa,\kappa_0,\beta)=0$.

We have thus shown that all solutions that satisfy $\frac{\partial (\Omega_\text{bh}+\Omega_\text{shell})}{\partial \sigma}\big|_M = 0$ also satisfy $\Psi = 0$.

\end{document}